\documentstyle[12pt]{article}
\begin{document}
\def\Tha{\Theta^{\alpha}}
\def\Thb{\Theta^{\beta}}
\def\da{{\dagger}}
\def\dda{{\ddagger}}
\def\tha{\theta^{\alpha}}
\def\thb{\theta^{\beta}}
\def\th+{\theta^+}
\def\th-{\theta^-}
\def\del{\partial}
\def\a{\alpha}
\def\e{\varepsilon}
\def\b{\beta}
\def\G{\Gamma}
\def\C{C_{\a\b}}
\def\cA{\cal{A}}
\def\os2{$OSp(2,2)$}
\def\su{$SU(2)$}
\def\be{\begin{equation}}
\def\ee{\end{equation}}
\def\bi{\bibitem}
\begin{titlepage}
\begin{flushright}
CERN-TH/95-195\\
UWThPh-20-1995\\
hep-th/9507074
\end{flushright}
\begin{center}
{\large\bf Field Theory on a Supersymmetric Lattice} \\
\vskip 1cm
{\bf H. Grosse\footnote{Part of Project No. P8916-PHY of the
`Fonds zur F\"orderung der wissenschaftlichen Forschung in
\"Ostereich'.}}\\
\vskip 0.3cm
{\it Institute for Theoretical Physics, University of Vienna
,}\\{\it Boltzmanngasse 5, A-1090 Vienna, Austria,}\\ \vskip 0.5cm
{\bf C. Klim\v c\'\i k} \\
\vskip 0.3cm
  {\it Theory Division CERN, CH-1211 Geneva 23,

Switzerland} \\
\vskip 0.5cm {\small and} \\
\vskip 0.5cm
 {\bf P. Pre\v snajder}\\
\vskip 0.3cm
 {\it Dept. of Theoretical Physics, Comenius
University,} \\ {\it Mlynsk\'a dolina F1, SK-84215 Bratislava,
 Slovakia} \\
\end{center}
\vskip 1cm
\begin{abstract}

A lattice-type regularization of the supersymmetric field theories
on a supersphere is constructed by approximating the ring of
scalar superfields
 by an integer-valued
sequence of finite dimensional rings of supermatrices and by
using the differencial calculus of non-commutative geometry.
 The regulated theory involves only finite number of
 degrees of freedom and is manifestly supersymmetric.
\end{abstract}
\vskip 0.5cm
\noindent CERN-TH/95-195\\
July 1995
\end{titlepage}

\section{\bf Introduction}

The idea that a fine structure of space-time should be
influenced by quantum gravity phenomena is certainly not
original  but so far there was a little success in giving it
more quantitative expression. String theory constitutes itself
probably the most promising avenue to a consistent theory of
quantum gravity it is therefore of obvious interest to study
the structure of spacetime from the point of view. Though
string theory incorporates a minimal lenght the physical
quantities computed in its framework reflect the symmetry
properties of continuous space-time. The situation is somewhat
analogous to ordinary quantum mechanics: though the phase
space acquires itself a cell-like structure its symmetries
remain intact, in general. In a sense the space-time possesses
the cell-like structure also in string theory e.g. the quantum
WZNW model for a compact group has as effective target,
perceived by a string center of mass, a truncated group
manifold or, in other words, a `manifold' with a cell-like
structure (see \cite{FrGaw}). Indeed, the zero-modes' subspace
of the full Hilbert space contains only the irreducible
representations of a spin lower than the level $k$. Because
this subspace describe the scalar excitations, it is clear
that high frequency (or spin) modes in an effective field
theory are absent. In this way string theory leads to the UV
finite behaviour of physical amplitudes as was probably
realized by several researchers in past (e.g.\cite{Kou}).

In our contribution we would like to initiate an investigation
of similar regularization in pure field theory context. That is we
wish to consider fields living on truncated compact manifolds,
endow them with dynamics and establish rules of their
quantization. Among advantages of such a development, there
would be not only the manifest preservation of all symmetries
of a theory but also an expected compatibility with quantum
gravity and string phenomena.
In some sense we shall construct a lattice-type of regularization
but the `lattice' will not approximate the underlying
spacetime (and hence the ring of functions on it) but directly the ring.
 As the starting point of our
treatment we choose a 2d field theory on a truncated two-sphere
\footnote{also referred to as ``fuzzy'', ``non-commutative''
or ``quantum'' sphere in literature \cite{Mad,Hop,Ber}.}.

The truncated sphere was extensively studied in past two
decades for various reasons. Apparently, the structure was
introduced by Berezin in 1975 \cite{Ber} who quantized the
(symplectic) volume two-form on the ordinary two-sphere.
He ended up with a series of possible quantizations
parametrized by the size of quantum cells. In 1982, Hoppe
\cite{Hop} investigated properties of spherical membranes. As
 a technical tool he introduced the truncation of high
 frequency excitations which effectively lead to the quantum
sphere. In 1991 the concept was reinvented  by Madore \cite{Mad}
(see also \cite{GroPre}). His motivation originated in the
so-called non-commutative geometry, i.e. the generalization
of the ordinary differential geometry to non-commutative
rings of `functions'. The truncated algebra of ordinary
functions is just the example of such a non-commutative ring.

For our purposes, we shall use the results of all those
previous works, however, we shall often put emphasis on
different aspects of formalism as comparing to the previous
investigations. Our main concern will consist in developing
basic differential and integral structures for non-commutative
sphere which are needed to define a classical (and quantum)
field dynamics. We shall require that the symmetries of the
undeformed theory are preserved in the non-commutative
deformation such as space-time supersymmetry, global isospin, local
(non)abelian gauge or chiral symmetry\footnote{An attempt
to formulate a field theory on the fuzzy sphere was published
in \cite{Mad,GroMad,GKP1}. However, the crucial concept of chirality
 was not studied there.} and,  obviously, that
the commutative limit should recover the standard formulation
of the dynamics of the field theory.

In many respects a canonical procedure for endowing
non-commutative rings with differential and integral calculus
is known for several years from basic studies of A. Connes
\cite{Con}. From his work it follows that geometrical
properties of a non-commutative manifold are encoded in
a fundamental triplet $(A,H,D)$ where $A$ is the representation
of a non-commutative algebra $\cal{A}$ of `functions' on the manifold in
 some Hilbert space. Elements of $A$ are linear operators
acting on $H$ in such a way that the multiplication of elements
of the `abstract' algebra $\cal{A}$ is represented by the
composition of the operators from $A$ which represent them.
$D$ is a self-adjoint operator (called the Dirac operator)
odd with respect to an appropriate grading\footnote{We ignore
in this paper aspects concerning the norms of the operators
from $A$ and commutators of the form $[D,A]$ because all
algebras we consider are finite-dimensional.} $H$ is
interpreted as a spinor bundle over the non-commutative
manifold and the action of the algebra $A$ on it makes
possible to define the action of a (truncated) gauge group
on spinors.

Noncommutative geometry has been already applied in theoretical physics
by providing the nice geometrical description of the standard model
action including the Higgs fields \cite{Con,CL}. The latter were
interpreted as the components of a noncommutative gauge connection.
Starting in this paper, we hope to provide another relevant application
of non-commutative geometry with the aim to understand the short
distance behaviour of field theory. We believe that non-commutative
geometry can provide powerful technical tools for performing new and
nontrivial relevant calculations.

In the present contribution, we construct the fundamental triplet
$(A,H,D)$
and use the construction for developing the supersymmetric regularization
of field theories. Though the uniqueness of $(A,H,D)$ for a given
fundamental algebra $\cal{A}$ is by no means guaranteed we give a highly
natural choice stemming from the following construction.
First we give a suitable description of spinors on the ordinary sphere
as components of a scalar superfield on a supersphere. Then we represent
the standard Dirac operator on the sphere in terms of the
superdifferential
generators of $OSp(2,1)$ algebra which is the supersymmetry superalgebra
of the supersphere. The standard Dirac operator on the sphere
turns out to be nothing
but the fermionic part of the Casimir of $OSp(2,1)$ written in the
superdifferential representation (the bosonic part is the standard
Laplace
operator on the sphere). Then we shall mimick the same construction for
the non-commutative sphere. We describe spinors on the non-commutative
sphere as the suitable components of a scalar superfield on a
non-commutative
supersphere. In other words, we perform the supergeometric
Berezin-like quantization
of the supersphere\footnote{Recently several papers have appeared dealing
with supergeometric quantization of the Poincar\'e disc
\cite{BKLR,Elg,ElgN}.} but in the language of  Madore.
The resulting quantized ring of scalar superfields will reveal a cell-like
structure of the non-commutative supersphere. The algebra $\cal{A}$ will
be the enveloping algebra of $OSp(2,1)$ in its irreducible representation
with a spin $j/2$. As $j\to\infty$ one recovers the standard ring of
 superscalar
functions on the supersphere. The quantized ring constitutes itself the
representation space of the adjoint action of $OSp(2,2)$ in the irreducible
representation with the $OSp(2,1)$ superspin $j/2$. We postulate that the
fermionic part of
the $OSp(2,1)$ Casimir in this adjoint representation is the Dirac operator
on the
non-commutative sphere. We shall find that it is selfadjoint and odd.
We shall compute its complete spectrum of eigenvalues and eigenfunctions
and find a striking similarity with the commutative case. Namely, the
non-commutative Dirac operator turns out simply to be a truncated commutative
one!\footnote{This suggests, in turn, that in the regulated field theory
one should avoid the problem of fermion doubling \cite{FerD}.} We then
construct both Weyl (chiral) and Majorana fermions.

The building of the supersymmetric theories requires even more
structure. We shall demonstrate that enlarging the superalgebra $OSp(2,1)$
to $OSp(2,2)$ the additional odd generators can be identified with the
supersymmetric covariant derivatives and the additional even generator
with the grading of the Dirac operator. All encountered representations
of $OSp(2,1)$ will turn out to be also the representations of $OSp(2,2)$.

In the following section (which does not contain original results) we
repeat the known construction of the standard non-commutative sphere
in a language suitable for SUSY generalization. In section 3 we give
the full account of the spectrum of the standard Dirac operator on the
commutative sphere. Though not the results themselves, but the (algebraic)
method of their derivation is probably new and very suitable for the later
non-commutative analysis. From the fourth section we present original
results. We start with the description of the (untruncated) Dirac operator
in terms of the fermionic part of the $OSp(2,1)$ Casimir acting on the ring
of superfields on the supersphere and  we quantize that ring.
Then we identify the Dirac operator on the non-commutative
sphere, give full account of its spectrum and describe the grading of the
non-commutative spinor bundle, completing thus the construction of the
fundamental triplet $(A,H,D)$. In section 5 we apply the developed
constructions in (supersymmetric) field theories. We shall construct
(super)symmetric action functionals of the deformed theories containing
only finite number of degrees of freedom.
We finish with conclusions and outlook concerning the construction
of a noncommutative de Rham complex,
a non-commutative gauge connection,
chiral symmetry, dynamics of gauge fields and
 construction of twisted bundles over the non-commutative sphere needed for
the description of `truncated' monopoles.

\section{\bf The non-commutative sphere}
\subsection{\bf The commutative warm-up}

A very convenient manifestly $SU(2)$ invariant description
of the ($L_2$-normed) algebra of functions $\cal{A}_{\infty}$
 on
 the ordinary sphere can be obtained by factorizing the
algebra $\cal{B}$ of analytic functions of three real
variables by its ideal $\cal{I}$, consisting of all functions
of a form $h(x^i)(\sum {x^i}^2-\rho^2)$. The scalar product on
$\cA_{\infty}$ is given by\footnote{The normalization ensures that the
norm of the unit element of  $\cal{A}_{\infty}$ is $1$.}
\be (f,g)_{\infty}\equiv{1\over 2\pi \rho}\int_{R^3}d^3 x^i \delta
({x^i}^2-\rho^2)f^*(x^i)g(x^i),\qquad f,g\in\cal{A}_{\infty}\ee
Here $f(x^i),g(x^i)\in\cal{B}$ are some representatives of $f$ and $g$.
The algebra $\cal{A}_{\infty}$ is obviously generated by
 functions\footnote{Speaking more precisely,
$x^i$ denote the corresponding equivalence classes in $\cal{B}$.}
 $x^i, i=1,2,3$
 which commute with each other under
the usual pointwise multiplication.
Their norms are given by
\be\vert\vert x^i\vert\vert^2_{\infty}={\rho^2\over 3}.\qquad\ee
 Consider the vector fields in $R^3$
generating \su \   rotations of $\cal{B}$. They are given by explicit
formulae
\be R_j=-i\epsilon_{jkl}x^k{\partial\over\partial x^l}\ee
and obey the \su \  Lie algebra commutation relations
\be [R_i,R_j]=i\epsilon_{ijk}R_k\ee
The action of $R_i$ on $\cal{B}$ leaves the ideal $\cal{I}$ invariant
hence it induces an action of \su \ on $\cal{A}_{\infty}$.
The generators $x^i\in\cal{A}_{\infty}$ form
a spin $1$ irreducible representation of \su \  algebra under the action
(hence they are linear combinations of the spherical functions with $l=1$).
They fulfil an obvious relation
\be {x^i}^2=\rho^2.\ee
 Higher powers of $x^i$
can be rearranged into irreducible multiplets corresponding to higher
 spins. For instance, the  multiplet of spin $l$ is conveniently
constructed subsequently applying the lowering operator $R_-\equiv R_1-iR_2$
on the highest weight vector ${x^+}^l$.
 It is well-known (cf.  any textbook on quantum mechanics)
that the
full decomposition of $\cal{A}_{\infty}$ into the irreducible representations
of \su \   is given by the infinite direct sum
\be {\cal{A}_{\infty}}=0+1+2+\dots,\ee
where the integers denote the spins of the representations.
\subsection{\bf The truncation of $\cal{A}_{\infty}$}

We define the family of non-commutative spheres ${\cal{A}}_j$ by furnishing
the truncated sum of the irreducible representations
\be {{\cal{A}}_j}=0+1+\dots+j,\ee
with an associative product and a scalar product
 which in the limit $j\to\infty$
give the standard products in ${\cal{A}}_{\infty}$.
To do this consider the space ${\cal{L}}(j/2,j/2)$ of linear operators
from the representation space of the irreducible representation with the
spin $j/2$ into itself. Clearly, \su \ algebra
acts on ${\cal{L}}(j/2,j/2)$ by the adjoint action. This `adjoint'
 representation
is reducible and the standard Clebsch-Gordan series for \su \  \cite{Lan}
gives
its decomposition
\be {\cal{L}}(j/2,j/2)=0+1+\dots +j\equiv{\cal{A}}_j.\ee
The scalar product on ${\cal{A}}_j$ is defined by\footnote{The normalization
ensures that the norm of the identity matrix is $1$.}
\be (f,g)_j\equiv{1\over j+1}{\rm Tr}(f^* g), \qquad f,g\in {\cal{A}}_j,
\ee
and the associative
product is defined as the standard composition of operators from
the space ${\cal{L}}(j/2,j/2)$.
Now we make more precise the notion of the commutative limits of the
scalar product and the associative product.
 There is a natural chain of the linear embeddings of the vector spaces
\be {{\cal{A}}_1}\hookrightarrow{{\cal{A}}_2}
\hookrightarrow\dots\hookrightarrow{{\cal{A}}_j}
\hookrightarrow\dots\hookrightarrow{{\cal{A}}_{\infty}}\qquad\ee
Any (normalized) element from  ${\cal A}_j$
of the form
\be c_{j,lp}R_-^p {X_j^+}^l \ee
is mapped in an (normalized)  element from ${\cal A}_k$ given by
\be c_{k,lp}R_-^p {X_k^+}^l.
\ee
Here
$X_{j}^{\a}$ are representatives of the \su \ generators in the irreducible
representation with spin $j/2$ ($X_{\infty}^{\a}\equiv x^{\a}$).
 They are normalized so that
\be [X_j^m,X_j^n]=
i{\rho\over\sqrt{{j\over 2}({j\over 2}+1)}}\epsilon_{mnp}X^p,\qquad\ee
and $c_{j(k),lp}$ are the (real) normalization coefficients given by the
 requirement
that the embedding conserves the norm. Note that ${X_{j}^+}^l$ are the
highest weight vectors in ${\cal A}_j$.
 Because the adjoint action of the \su \
algebra is hermitian for arbitrary  ${{\cal{A}}_j}$
(as it can be easily seen from the definitions of the scalar products (1),(9))
the embeddings are in fact isometric.
Indeed, the scalar product of the eigenvectors
of the hermitian operator vanishes if the corresponding
eigenvalues are different. Obviously different $l$'s give different
eigenvalues of the (hermitian) adjoint Casimir. The
commutative limit of the associative product is more
involved, however\footnote{The nice establishment of the correct commutative
limit of the product was given in \cite{GroPre} using the coherent states
for \su .}.
 Clearly, the embeddings cannot be (and should not be) the
homomorphisms of the associative products! For instance the product of two
elements from ${{\cal{A}}_j}$ with the maximal spin $j$
 has again  a maximal spin $j$ because it is from ${{\cal{A}}_j}$ but could
have a spin $2j$ component
if the product is taken in a sufficiently larger algebra
 ${{\cal{A}}_k}$.

Consider more closely the behaviour of the product as the function
of $k$. According (10),
 arbitrary two elements $f,g$ of ${{\cal{A}}_j}$  can be canonically
considered as the elements of ${{\cal{A}}_k}$ for whatever $k>l$ (including
$k=\infty$). Their product in every  ${{\cal{A}}_k}$ can also be embedded
in ${\cal{A}}_{\infty}$. Denote the corresponding element of
${\cal{A}}_{\infty}$ as $(fg)_k$.
 We shall argue that
\be \displaystyle{\lim_{k\to\infty}{(fg)_k}}=fg\qquad\ee
where $fg$ is the standard commutative pointwise multiplication in
${\cal{A}}_{\infty}$ .

Before plunging into proof of this statement we try to formulate
its meaning more
`physically'. It is not true that the algebra ${{\cal{A}}_j}$ tends to be
commutative for large $j$ (as the matrix algebra it, in fact, cannot.)
What is the case that for large $j$ the elements with much lower spins than
$j$ almost commute. In the field theory language: long distance limit
corresponds to the standard commutative theory but for short distances the
structure is truly non-commutative.
This non-commutativeness, however, preserves the symmetry
of the space-time. The algebra ${{\cal{A}}_j}$ is finite-dimensional with the
dimension being $(j+1)^2$. That means that the sphere is effectively divided
in $(j+1)^2$ cells of an average area ${4\pi\rho^2\over(j+1)^2}$. A
theory based on the non-commutative ring ${{\cal{A}}_j}$ has, therefore,\
a minimal lenght ${2\rho\over j+1}$ incorporated.

Now it is easy to prove (14). Actually because of relation (13), which
ensures the commutativity of the limit, it is enough to show that the
normalization coefficients $c_{j,lp}$  defined in (11,12) have the property

\be \displaystyle{\lim_{k\to\infty}{c_{k,lp}}}=c_{\infty,lp}.\ee
Due to the rotational invariance of the inner products in all
${\cal A}_k (k=1,\dots,\infty)$ it is enough to demonstrate it just
for the highest weight element ${X_k^+}^l$. Then
\be  c_{k,l0}^{-2}=({X_k^+}^l,{X_k^+}^l)_k=\rho^{2l}{(2l)!!\over (2l+1)!!}
{(k+l+1)!\over (k+1)(k)^l(k+2)^l(k-l)!}\ee
The last equality follows from a formula derived in \cite{Pru}
(p. 618, Eq. (36)).

The relation (15) then obviously holds since the last fraction
tends to $1$ and it can be simply computed from (1) that
\be c_{\infty,l0}^{-2}=\rho^{2l}{(2l)!!\over (2l+1)!!}.\ee
 Note that the generators $X_k^i$ are themselves normalized as
\be (X_k^i,X_k^i)_k={\rho^2\over 3}\ee
and the standard relation defining the surface $S_2$ holds
in the non-commutative case
 \be {X_k^i}^2=\rho^2.\ee
We observe from (2) and (18)
that for every $j$
$X_j^i\in {{\cal{A}}_j}$ are embedded in ${{\cal{A}}_{\infty}}$
as just the standard commutative generators $x^i$
and
in ${{\cal{A}}_k},k>j$ as $X_k^i\in {{\cal{A}}_k}$ .  The notation is
therefore
justified and in what follows we shall often write just $X^i$ in the
non-commutative case and $x^i$ in the commutative one.

\section{\bf The Dirac operator on $S_2$ and its spectrum}

The construction of the spinor bundle\footnote{We have in mind the trivial
bundle, twists by $U(1)$ bundles needed for the inclusion of monopoles
will be considered in a forthcoming paper.}
 over $S_2$ is standard part of any
textbook of quantum field theory (e.g. see \cite{IZ}) though, perhaps, it
is not stressed explicitly. Also the spectrum of the Dirac operator acting
on this bundle is known in that context, the eigenfunctions are nothing but
the so-called spinorial harmonics \cite{IZ}. We present the manifestly
rotation invariant description of the
spectrum in the spirit of the previous section.

Consider the trivial spinor bundle $S_B$ over ${\bf R}^3$. Its sections are
ordinary
quantum mechanical two-component spinorial wave-functions of the form
\be \left(\matrix{\Psi_+\cr\Psi_-}\right),\qquad
\Psi_+,\Psi_-\in{\cal{B}}. \ee  The action of the \su \ algebra is described
by the generators
\be J_i\equiv R_i+{1\over 2}\sigma_i,\ee
where $\sigma_i$ are the standard Pauli matrices.
Hence, $S_B$ is the representation space of some (reducible) representation
of \su \ .
Now $R_3$ can be viewed as the fibration of $S_2$ by the half-lines in $R_3$
starting in its centre. The position of a point on the fiber we measure
by the radial coordinate $r$. The subbundle $S_{A_{\infty}}$ of the sections
of $S_B$ independent on the fiber coordinate $r$ can be interpreted as
the spinor bundle over the base manifold $S_2$ of the fibration. Clearly,
$S_{A_{\infty}}$ is the \su \  subrepresentation of $S_B$. The decomposition
of $S_{A_{\infty}}$ into irreducible representation follows from the standard
Clebsch-Gordan series  \cite{Lan} for the tensor product of the representations
${\cal{A}}_{\infty}$ and $1/2$
\be S_{A_{\infty}}=2(1/2+3/2+5/2+\dots).\ee
Here the factor $2$ in front of the bracket means that each representation
in the bracket occurs  in the direct sum twice. This doubling may be
interpreted as the sum of the left and right chiral spinor bundles.
We shall argue that the standard Dirac operator corresponding to the round
metric on $S_2$ can be written solely in terms of the \su \  generators as
follows\footnote{The same formula was already given in \cite{Jay,GP}. We
give the different evidence of its validity, however.}
\be D={1\over\rho}(\sigma_i R_i+1).\ee
Here $\rho$ is the radius of the sphere. This operator is self-adjoint with
respect to the scalar product on  $S_{A_{\infty}}$ given by
\be (\Psi,\Xi)\equiv {1\over 2\pi\rho}\int d^3x^i\delta({x^i}^2-\rho^2)
(\Psi_+^*\Xi_+ +\Psi_-^*\Xi_-),\qquad\Psi,\Xi\in S_{A_{\infty}}.\ee

The easy way of deriving (23) consists in comparing a three dimensional
flat Dirac operator $D_3$ on $S_B$ written in the spherical coordinates with
the two dimensional round Dirac operator $D_2$ on the sphere in the same
 coordinates.
Due to the rotational invariance the choice of a coordinate chart is irrelevant
and we may proceed by choosing (and fixing) the poles of the sphere.
The  Dirac operator D in arbitrary coordinates in a general (curved)
Riemannian manifold is given by
\be D_2=-i\gamma^a e_a^{\mu}(\partial_{\mu}+{1\over 4}\omega_{\mu ab}[\gamma^a,
\gamma^b]),\ee
where $\gamma^a$ are generators of the flat Clifford algebra
 \be \{\gamma^a,
\gamma^b\}=2\delta^{ab},\qquad {\gamma^a}^2=1,\qquad {\gamma^a}^{\dagger}=
{\gamma^a},\ee
 $e_a^{\mu}$ is the vielbein and $\omega_{\mu ab}$ the spin connection defined
by
\be \partial_{\mu}e^a_{\nu}-\Gamma^{\lambda}_{\mu\nu} e^a_{\lambda}+
\omega^{~a}_{\mu ~b}e^b_{\nu}=0.\ee
For $S_2$ in the spherical coordinates
\be e_1^{\theta}={1\over\rho},
\qquad e_2^{\phi}={1\over\rho\sin{\theta}},\qquad\omega_{\phi 12}=
-\omega_{\phi 21}=-\cos{\theta}.
\ee
All remaining components of the vielbein and the connection vanish.
For $R^3$ in the spherical coordinates
\be e_1^{\theta}={1\over r},
\qquad e_2^{\phi}={1\over r\sin{\theta}},\qquad e_3^r=1\ee
and
\be \omega_{\phi 21}=
-\omega_{\phi 12}=\cos{\theta},
\qquad\omega_{\phi 23}=
-\omega_{\phi 32}=\sin{\theta},\qquad\omega_{\theta 13}=
-\omega_{\theta 31}=1.\ee
Thus
\be D_2=-i\gamma^1{1\over \rho}(\partial_{\theta}+{1\over 2}{\rm ctg}\theta)
-i\gamma^2{1\over \rho\sin{\theta}}\partial_{\phi}.\ee
and
\be D_3=-i\gamma^1{1\over r}(\partial_{\theta}+{1\over 2}{\rm ctg}\theta)
-i\gamma^2{1\over r\sin{\theta}}\partial_{\phi}+
-i\gamma^3(\partial_r+{1\over r}).
\ee
We observe a simple relation between $D_3$ restricted on $S_{A_{\infty}}$
and $D_2$ namely
\be -i\gamma^3 D_3\vert_{restr.} +1/\rho = D_2.\ee
(note that $-i\gamma^3\gamma^a, a=1,2$ fulfil the defining relations of
the Clifford algebra (26)).

$D_3$ can be expressed also in the flat coordinates in $R^3$
\be D_3=-i\sigma_i \partial_i,\ee
where $\sigma_i$ are the Pauli matrices which also generate the Clifford
algebra (26).
A simple algebra gives
\be D_3=\big({\sigma_k x_k\over r}\big)^2D_3
=-i\big({\sigma_k x_k\over r}\big)\big({x_i\over r}\partial_i -{1\over r}
\sigma_i R_i\big).\ee
Because ${x_i\over r}\partial_i = \partial_r$
 and the vector fields $R_i$
have no radial component it follows from (32) and (35) that
\be \gamma_3=\big({\sigma_k x_k\over r}\big).\ee
Inserting $\gamma_3$ from (36) and $D_3$ from (35) into Eq.(33) we get the
\su \  covariant
form (23) of the round Dirac operator on $S_2$.

The spectrum of $D_2$ readily follows from the group representation
considerations. Consider a (normalized) spinor
\be {\Theta^+\over \rho}=\left(\matrix{1\cr 0}\right).\ee
It is obviously the eigenvector of $D_2$ with an eigenvalue 1. Moreover it
is the highest weight state of one of the spin $1/2$ representations
in the decomposition (22) as it can be directly checked using the generators
$J_i$ from (21). Indeed
\be J_+\Theta^+=0,\qquad J_iJ_i~\Theta^+=3/4.\ee
The construction of the other
(normalized) highest weight states in the irreducible
representations with the higher spins is obvious. They are given by
\be \Psi_{l,h.w.}=\rho^{-l-1}\sqrt{{(2l+1)!!\over (2l)!!}} {x^+}^l\Theta^+ .
\ee
Here  $l$
is the spin of the irreducible representation. A direct computation shows

\be D_2\Psi_{l,h.w.}=(l+1)\Psi_{l,h.w.}.\ee
Due to the rotational invariance of $D_2$ the other eigenvectors within
the irreducible representation are obtained by the action of the lowering
generator $J_-$, i.e.
\be \Psi_{l,m}=\rho^{-l-1}\sqrt{{(2l+1-m)!\over (2l+1)! m!}
{(2l+1)!!\over (2l)!!}}J_-^m {x^+}^l \Theta^+.
\ee
The eigenvalue corresponding to the eigenvector $\Psi_{l,m},~~m=0,\dots,
2l$ is obviously $l+1$. So far we have constructed only one branch
of the spectrum. However, due to an obvious relation
\be D_2\gamma^3+\gamma^3 D_2=0\ee
also spinors $\gamma^3\Psi_{l,m}$ are the eigenvectors of $D_2$ with
the eigenvalues $-(l+1)$. In this way we found the complete spectrum because
all eigenvectors $\Psi_{l,m}$ and $\gamma^3\Psi_{l,m}$ form the basis of the
spinor bundle  $S_{A_{\infty}}$.
\section{\bf Non-commutative supersphere}

Having in mind the  goal of constructing a non-commutative spinor
bundle, we have to look for a language to describe the commutative case
which would be best suited for performing the non-commutative deformation.
We shall argue that the very structure to be exploited is \os2 ~ superalgebra
which
is somewhat hidden in the presentation given in the previous section.  We
shall proceed conceptually as follows: The
non-commutative sphere, described in section 2, emerged naturally from the
quantization of the algebra of the scalar fields on the ordinary sphere.
Hence, it is natural to expect that the quantization of the supersphere would
give a deformed ring of the scalar superfields on the supersphere. Those
superfields contain as their components the ordinary fermion fields on the
sphere, therefore the deformation of the algebra of the superfield should
give ( and it does give) the non-commutative spinor bundle on the
non-commutative sphere, i.e. the structure we are looking for.
\subsection{\bf (Super)commutative supersphere}

Consider a three-dimensional superspace ${\bf SR^3}$ with coordinates
$x^i,\theta^{\alpha}$;
the super-coordinates are the \su ~ Majorana spinors. Consider an algebra
${\cal SB}$ of
 analytic functions on the superspace with the Grassmann coefficients in front
of the odd monomials in $\theta$. ${\cal SB}$ can  be factorized by its ideal
${\cal SI}$,
consisting of all functions of a form $h(x^i,\tha)(\sum {x^i}^2+
C_{\a\b}\tha\thb -\rho^2)$. Here
\be C=i\sigma^2.\ee
 We refer to the quotient ${\cal SA}_{\infty}$ as to the algebra
of superfields on the supersphere. An \os2 ~invariant
 inner product of two elements $\Phi_1,\Phi_2$ of
 ${\cal SA}_{\infty}$ is given by\footnote{The normalization ensures that
the norm of the unit
element of ${\cal SA}_{\infty}$ is 1. The inner product is  supersymmetric
but it is not positive definite. However, such a property of the product is
not needed for our purposes.}
\be(\Phi_1,\Phi_2)_{\infty}\equiv{\rho\over 2\pi}\int_{R^3}d^3 x^i
d\theta^+ d\theta^-\delta
({x^i}^2+C_{\a\b}\tha\thb -\rho^2)\Phi_1^{\ddagger}(x^i,\tha)
\Phi_2(x^i,\tha),\ee
Here $\Phi_1(x^i,\tha),\Phi_2(x^i,\tha)\in{\cal SB}$ are some representatives
of $\Phi_1$ and $\Phi_2$ and the (graded) involution \cite{SNR2,SNR} is
defined
by
\be {\theta^+}^\dda=\theta^-,~{\theta^-}^\dda=-\theta^+,~
{}~(AB)^\dda=(-1)^{degA~deg B}B^\dda A^\dda .\ee The algebra
${\cal SA}_{\infty}$ is obviously generated by (the equivalence classes)
$x^i~ (i=1,2,3)$ and $\tha~ (\a=+,-)$ which (anti)commute
with each other under the usual pointwise multiplication, i.e.
\be x^i x^j-x^j x^i=x^i\tha-\tha x^i=\tha\thb+\thb\tha=0.\ee Their norms are
given by
\be \vert\vert x^i\vert\vert^2_{\infty}=\vert\vert \tha\vert\vert^2_{\infty}=
\rho^2.\ee
Consider the vector fields in ${\bf SR^3}$ generating \os2 ~superrotations of
${\cal SB}$. They are given by explicit formulae

\be v_+=-{1\over 2}\big(x^3\del_{\theta^-}-(x^1+ix^2)\del_{\theta^+}\big)+
{1\over 2}\big(-\theta^+\del_{x^3}-\theta^-(\del_{x^1}+i\del_{x^2})\big),\ee
\be v_-=-{1\over 2}\big(x^3\del_{\theta^+}+(x^1-ix^2)\del_{\theta^-}\big)+
{1\over 2}\big(\theta^-\del_{x^3}-\theta^+(\del_{x^1}-i\del_{x^2})\big),\ee
\be d_+=-{1\over 2}r(1+{2\over r^2}\theta^+ \th-)\del_- +{\th-\over 2r}R_+
-{\theta^+ \over 2r}(x^i\del_i-R_3),\ee
\be d_-={1\over 2}r(1+{2\over r^2}\theta^+ \th-)\del_+ +{\theta^+ \over 2r}R_-
-{\th-\over 2r}(x^i\del_i+R_3)\ee
\be \Gamma_{\infty}=({\theta^+ x^3\over r}+{\th- x^+\over r})\del_+ +
({\theta^+ x^-\over r}-{\th- x^3\over r})
\del_-  \equiv 2(\th- v_+ -\theta^+ v_-).\ee
\be r_+=x^3(\del_{x^1}+i\del_{x^2})-(x^1+ix^2)\del_{x^3}+\theta^+
\del_{\theta^-},\ee
\be r_-=-x^3(\del_{x^1}-i\del_{x^2})+(x^1-ix^2)\del_{x^3}+
\theta^-\del_{\theta^+},\ee
\be r_3=-ix^1\del_{x^2}+ix^2\del_{x^1}+{1\over 2}
(\theta^+\del_{\theta^+}-\theta^-\del_{\theta^-})\ee
and they obey the \os2 ~ Lie superalgebra graded commutation relations
\cite{ElgN,SNR}
\be [r_3,r_{\pm}]=\pm r_{\pm},\qquad [r_+,r_-]=2r_3,\ee
\be [r_3,v_{\pm}]=\pm{1\over 2}v_{\pm},\qquad [r_{\pm},v_{\pm}]=0,
\qquad [r_{\pm},v_{\mp}]=v_{\pm},\ee
\be\{v_{\pm},v_{\pm}\}=\pm{1\over 2}r_{\pm},\qquad \{v_{\pm},v_{\mp}\}=
-{1\over 2}r_3.\ee
\be [\G_{\infty},v_{\pm}]=d_{\pm},\qquad[\G_{\infty},d_{\pm}]=v_{\pm},
\qquad [\G_{\infty},r_i]=0,\ee
\be [r_3,d_{\pm}]=\pm {1\over 2}d_{\pm},\quad[r_{\pm},d_{\pm}]=0,
\quad [r_{\pm},d_{\mp}]=d_{\pm},\ee
\be \{d_{\pm},v_{\pm}\}=0,\qquad \{d_{\pm},v_{\mp}\}=
\pm{1\over 4}\G_{\infty},\ee
\be \{d_{\pm},d_{\pm}\}=\mp{1\over 2}r_{\pm},\qquad
\{d_{\pm},d_{\mp}\}={1\over 2}r_3.\ee
 Note, that all  introduced generators
do annihilate the quadratic form ${x^i}^2+C_{\a\b}\tha\thb$ hence
they induce the action of $OSp(2,2)$ on
${\cal SA}_{\infty}$\footnote{The appearance of r in Eqs.(50-52) may seem
awful because we have considered the ring of superanalytic functions
on $SR_3$. However, this is only a formal drawback, which can be
cured by a completion of the space of superanalytic functions with
respect to an appropriate inner product. In fact, we need not even
do that for our purposes because the terms involving  $r$ become
anyway harmless after the factorization by the ideal $SI$.}
{}.

In order to demonstrate the \os2 ~invariance of the inner product (44)
we have to settle the properties of the \os2 ~ generators
with respect to the graded involution. It holds
\be (\Phi_1, r_i\Phi_2)_{\infty}=(r_i\Phi_1,\Phi_2)_{\infty}\ee
\be (\Phi_1,v_{\mp}\Phi_2)_{\infty}=
\pm(v_{\pm}\Phi_1,\Phi_2)_{\infty}.\ee
\be (\Phi_1,d_{\mp}\Phi_2)_{\infty}=
\mp(d_{\pm}\Phi_1,\Phi_2)_{\infty}.\ee
\be (\Phi_1,\G_{\infty}\Phi_2)_{\infty}=
(\G_{\infty}\Phi_1,\Phi_2)_{\infty}.\ee
Consider now the variation of a superfield $\Phi$
\be \delta\Phi=i(\e_+ v_+ +\e_- v_-)\Phi,\ee
where $\e_{\a}$ is a constant Grassmann Majorana spinor, i.e.
\be \e^{\dda}_+=\e_-,\quad \e^{\dda}_-=-\e_+\ee
and, much in the same manner, a variation
\be \delta\Phi=i(\e_- d_+ +\e_+ d_-)\Phi.\ee
Using the relations (63-66) it is straightforward to observe the invariance
of the inner product with respect to the defined variations.

 As it is well known \cite{SNR} the typical irreducible representations of
\os2 ~ consist of quadruplets of the \su ~
irreducible representations $j\oplus j-{1\over 2}\oplus j-{1\over 2}\oplus
j-1$. The number $j$ is an integer
or a half-integer and it is referred to as the \os2 ~superspin. The generators
$x^i,\tha\in {\cal SA}_{\infty}$ together with
\be {1\over \rho^2}(\theta^+ x^3+\theta^- x^+),\qquad j={1\over 2},~j_3=
{1\over 2},\ee
\be {1\over \rho^2}(\theta^+ x^- -\theta^- x^3),
\qquad j={1\over 2},~j_3=-{1\over 2},\ee
\be 1+{1\over \rho^2}\theta^+\theta^-,\qquad j=0,~j_3=0.\ee
indeed form the (typical) superspin 1 irreducible representation
of \os2 ~algebra under the action of the vector fields (48-55).
The  numbers $j,j_3$ in (70-72) correspond to the total \su ~spin and its third
component.
The supermultiplet with the superspin $1$ can be conveniently constructed
applying subsequently the lowering operators $v_-$ and $d_-$ on the highest
weight vector $x^+$.
 Supermultiplets with higher superspins can be obtained
in the same way starting with the highest weight vectors ${x^+}^l$.
Thus
 the full decomposition of ${\cal SA}_{\infty}$ into the irreducible
representations of \os2 ~ can be written as  the infinite
direct sum
\be {\cal SA}_{\infty}=0+1+2+\dots,\ee
where the integers  denote the \os2 ~superspins of the
representations\footnote{The  `baryon' number of those representations,
in the sense of Ref.\cite{SNR},
is zero.}.
 From the point of view of the \su ~representations,
the algebra of the superfields consists of two copies of ${\cal A}_{\infty}$
and the spinor bundle ${1\over 2}\otimes {\cal A}_{\infty}$ (see Eq. (22))
 Note that the generators of ${\cal SA}_{\infty}$
fulfil the obvious relation
\be {x^i}^2+C_{\a\b}\tha\thb =\rho^2.\ee

The big algebra  ${\cal SB}$ has a natural grading as the vector space, given
by the parity of the total power of the Grassmann coordinates $\tha$. Because
we factorized over the quadratic surface in the superspace, this grading
induces the grading in  ${\cal SA}_{\infty}$. It is easy to see that the odd
elements of  ${\cal SA}_{\infty}$ with respect to this grading can be
identified with the fermion fields
on the sphere. Indeed, they  can be written as
\be \Psi=\Psi_{\a}(x^i){\tha\over\rho},\ee
where the (Grassmann) components $\Psi_{\a}$ belong to
${\cal A}_{\infty}$\footnote{
The factorization by the relation $\sum {x^i}^2 -\rho^2=0$ and the relation
$\sum {x^i}^2+C_{\a\b}\tha\thb -\rho^2=0$ is effectively the same in this
case because the term quadratic in $\theta $ is killed
upon the multiplication by another $\theta $ in Eq. (75).}. But this is
the  standard spinor
bundle on the sphere
\be \left(\matrix{\Psi_+(x^i)\cr\Psi_-(x^i)}\right),\ee
 described in section 3. The scalar product on the bundle is inherited
from the inner product (44)
\be (\Psi,\Xi)\equiv {\rho \over 2\pi}\int d^3x^i
\delta({x^i}^2-\rho^2)d\theta^+
d\theta^-
\Psi^\dda\Xi,\ee
and (up to a sign) it coincides with the scalar product (24). The Pauli
matrices, as
the operators acting on the
two-component spinors, can be expressed in the superfield formalism as follows
\be \sigma^3=\theta^+\del_{\theta^+} -\theta^-\del_{\theta^-},
\qquad \sigma^{\pm}=2\theta^{\pm}\del_{\theta^{\mp}}.\ee
In what follows we shall refer to the odd (even) elements with respect to
the described grading as to the fermionic (bosonic) superfields in order to
make a difference with the even and odd superfields in the standard
(Grassmann) sense.

The $OSP(2,1)$ superalgebra generated by $r_i,v_{\pm}$ has a quadratic
Casimir
\be K_2=(r_3^2+{1\over 2}\{r_+,r_-\})+(v_+v_- -v_-v_+)\equiv B_2 +F_2.\ee
Using Eqs. (78),
it is easy now to check  that the fermionic part $F_2$ of the Casimir is
directly related to the Dirac operator (23)
\be \rho D=\sigma^i R_i+1=
2F_2-{1\over 2}=2(v_+v_- -v_-v_+)-{1\over 2}.\ee
The grading $\gamma^3$ of the Dirac operator is just the \os2 ~generator
$\G_{\infty}$. Its eigenfuctions are obviously the Weyl spinors. A Majorana
spinors are given by the restriction
\be \psi_+^{\dda}=\psi_-,\qquad \psi_-^{\dda}=-\psi_+\ee
which can be easily derived from the reality condition on the superfield
$\Phi$.

\subsection{\bf The truncation of ${\cal SA}_{\infty}$}
We define the family of non-commutative superspheres ${\cal{SA}}_j$ by
furnishing
the truncated sum of the irreducible representations of \os2 ~
\be {{\cal{SA}}_j}=0+1+\dots+j,\qquad j\in{\bf Z}\ee
with an associative product and an inner product
 which in the limit $j\to\infty$
give the standard products in ${\cal{SA}}_{\infty}$.
In order to do this consider the space ${\cal{L}}(j/2,j/2)$ of linear
operators
from the representation space of the $OSp(2,1)$ ~irreducible representation
with the
$OSp(2,1)$ superspin $j/2$ into itself. (Note that the $OSp(2,1)$
irreducible representation
 with the $OSp(2,1)$ superspin $j$ has the \su ~content $j
\oplus j-{1\over 2}$ \cite{SNR}).  The action of the superalgebra
\os2~itself on
${\cal{L}}(j/2,j/2)$\footnote{The so-called non-typical irreducible
representation
of \os2 ~ \cite{SNR,Mar}  is in the same time also the $OSp(2,1)$
irreducible representation with the
$OSp(2,1)$ superspin $j/2$.} is described by operators $R_i,V_{\a},D_{\a},
\gamma\in{\cal{L}}(j/2,j/2)$ given by \cite{PR}
\be R_i=
\left(\matrix{R_i^{j\over 2}&0\cr 0&R_i^{{j\over 2}-{1\over 2}}}\right),
\qquad \gamma=\left(\matrix{-j~Id&0\cr 0& -(j+1)Id}\right).\ee
\be
 V_{\a}=\left(\matrix{0&V_{\a}^{{j\over 2},{j\over 2}-{1\over 2}}\cr
V_{\a}^{{j\over 2}-{1\over 2},{j\over 2}}&0}\right),
\qquad D_{\a}=\left(\matrix{0&V_{\a}^{{j\over 2},{j\over 2}-{1\over 2}}\cr -
V_{\a}^{{j\over 2}-{1\over 2},{j\over 2}}&0}\right)
,\ee

where
\be \langle l,l_3+1\vert R_+^l\vert l,l_3\rangle=\sqrt{(l-l_3)(l+l_3+1)},\ee
\be \langle l,l_3-1\vert R_-^l\vert l,l_3\rangle=\sqrt{(l+l_3)(l-l_3+1)},\ee
\be \langle l,l_3\vert R_3^l\vert l,l_3\rangle=l_3,\ee
\be \langle l_3+{1\over 2} \vert V_+^{{j\over 2},{j\over 2}-{1\over 2}}
\vert l_3\rangle=-{1\over 2}\sqrt{{j\over 2}+l_3+{1\over 2}},\ee
\be \langle l_3-{1\over 2} \vert V_-^{{j\over 2},{j\over 2}-{1\over 2}}\vert
l_3\rangle=-{1\over 2}\sqrt{{j\over 2}-l_3+{1\over 2}},\ee
\be \langle l_3+{1\over 2} \vert V_+^{{j\over 2}-{1\over 2},{j\over 2}}\vert
l_3\rangle=-{1\over 2}\sqrt{{j\over 2}-l_3},\ee
\be \langle l_3-{1\over 2} \vert V_-^{{j\over 2}-{1\over 2},{j\over 2}}\vert
l_3\rangle={1\over 2}\sqrt{{j\over 2}+l_3}.\ee
 Every  $\Phi\in{\cal{L}}(j/2,j/2)$ can be written as a matrix
\be \Phi=\left(\matrix{\phi_R &\psi_R\cr \psi_L &\phi_L}\right),\ee
where $\phi_R$ and $\phi_L$ are square $(j+1)\times (j+1)$ and $j\times j$
matrices respectively and $\psi_R$ and $\psi_L$ are respectively rectangular
$(j+1)\times j$ and $j\times(j+1)$ matrices. The meaning of the indices $R$
and $L$ will become clear in the next subsection. A fermionic  element is
given by
a supermatrix with vanishing diagonal blocks and a bosonic element by one
with vanishing off-diagonal blocks.
Clearly, \os2 \ superalgebra
acts on ${\cal{L}}(j/2,j/2)$ by the superadjoint action
\be {\cal R}_i\Phi\equiv [R_i,\Phi],\qquad \Gamma\Phi\equiv [\gamma,\Phi].\ee
\be {\cal V}_{\a}\Phi_{even}\equiv [V_{\a},\Phi_{even}],
\qquad {\cal V}_{\a}\Phi_{odd}\equiv \{V_{\a},\Phi_{odd}\}.\ee
\be {\cal D}_{\a}\Phi_{even}\equiv [D_{\a},\Phi_{even}],
\qquad {\cal D}_{\a}\Phi_{odd}\equiv \{D_{\a},\Phi_{odd}\}.\ee

 This `superadjoint'
 representation
is reducible and, in the spirit of Ref.\cite{SNR, Mar}, it is easy to work
out its decomposition into \os2 ~ irreducible representations
\be {\cal{L}}(j/2,j/2)=0+1 +\dots +j.\ee
 The associative product in ${\cal{L}}(j/2,j/2)$ is defined as the
composition of operators  and
the \os2 ~invariant inner product on ${\cal{L}}(j/2,j/2)$ is defined
by\footnote{The normalization
ensures that the norm of the identity matrix is $1$.}
\be (\Phi_1,\Phi_2)_j\equiv {\rm STr}(\Phi_1^\dda,\Phi_2),
\qquad \Phi_1,\Phi_2\in {\cal{L}}(j/2,j/2).
\ee
Here $STr$ is the supertrace and $\dda$ is the graded involution.
Although
these concepts are quite standard in the literature it is instructive to work
 out
their content in our concrete example.  The supertrace is defined as usual
\be STr\Phi\equiv Tr\phi_R -Tr\phi_L\ee
and the graded involution as \cite{SNR2}
\be \Phi^\dda\equiv
\left(\matrix{\phi_R^\da &\mp\psi_L^\da\cr\pm\psi_R^\da &\phi_L^\da}\right).\ee
$\da$ means the standard hermitian conjugation of a matrix and the upper
(lower) sign refers to the case when the entries consists of odd (even)
elements
of a Grassmann algebra.
Note that
\be R_i^\dda=R_i,\qquad V_+^\dda=V_-,\quad V_-^\dda=-V_+.\ee
Now we  identify   ${\cal SA}_j$ with even  elements of ${\cal{L}}(j/2,j/2)$
which means that the entries of the (off)-diagonal matrices are
(anti)-commuting variables.
 This correspond to the similar
requirement in the untruncated case because in the truncated case the spinors
form the off-diagonal part of the superfield.

 We can demonstrate the
\os2 ~invariance of the inner product (97) again by settling the  properties
of the \os2 ~generators with respect to the graded involution (99).
They read
\be (\Phi_1,{\cal R}_i\Phi_2)_j=
({\cal R}_i\Phi_1,\Phi_2)_j.\ee
\be (\Phi_1,{\cal V}_{\mp}\Phi_2)_j=
\pm({\cal V}_{\pm}\Phi_1,\Phi_2)_j.\ee
\be (\Phi_1,{\cal D}_{\mp}\Phi_2)_j=
\mp({\cal D}_{\pm}\Phi_1,\Phi_2)_j.\ee
\be (\Phi_1,\G\Phi_2)_j=
(\G\Phi_1,\Phi_2)_j.\ee
Consider now the variation of a superfield $\Phi$
\be \delta\Phi=i(\epsilon_+ {\cal V}_+ +\epsilon_- {\cal V}_-)\Phi,\ee
where $\epsilon_{\a}$ is given by
\be \epsilon_{\a}=\left(\matrix{\e_{\a}&0\cr 0&-\e_{\a}}\right)\ee
and $\e_{\a}$ are the usual Grassmann variables with the involution properties
\be \e^{\dda}_+=\e_-,\quad \e^{\dda}_-=-\e_+ . \ee
Much in the same manner, consider also  a variation
\be \delta\Phi=i(\epsilon_- {\cal D}_+ +\epsilon_+ {\cal D}_-)\Phi.\ee
Using the relations (101-104) it is straightforward to observe the invariance
of the inner product with respect to the defined variations.
Note that $\epsilon_{\a}$ do anticommute with $D_{\a}$ and $V_{\a}$ as they
should.

We can choose a basis in  ${\cal{SA}}_j$ formed by eigenvectors of the
Hermitian operators
\be{\cal Q}^2\equiv{\cal R}_i^2+C_{\a\b}{\cal V}_{\a}{\cal V}_{\b},\ee
${\cal R}_i^2$ and ${\cal R}_3$. The spectrum of (the $OSp(2,1)$ Casimir)
${\cal Q}^2$ consists of
numbers $q(q+1/2)$ where the $OSp(2,1)$ superspin $q$ runs over all integers
and
half-integers from $0$ to $j$ \cite{PR}; the remaining two
operators have the standard
spectra known in the \su ~context.

Now we make more precise the notion of the commutative limits of the
inner product and the associative product.
 There is a natural chain of the linear embeddings of the vector spaces
\be
{{\cal{SA}}_1}\hookrightarrow{{\cal{SA}}_2}\hookrightarrow\dots
\hookrightarrow {{\cal{SA}}_j}
\hookrightarrow\dots\hookrightarrow{{\cal{SA}}_{\infty}}\qquad\ee
Any (normalized) element from ${\cal{SA}}_j$ of a form
\be s_{j,lpq}{{\cal V}_-}^p {{\cal D}_-}^q {X_j^+}^l \ee
is mapped into an element from ${\cal{SA}}_k$ of the form
\be s_{k,lpq}{{\cal V}_-}^p {{\cal D}_-}^q {X_k^+}^l.\ee
Here $X_j^i$ (and $\Theta_j^{\pm}\equiv -{\cal V}_{\mp} X_j^{\pm}$) are the
representatives
of the $OSp(2,1)$ generators in the $OSp(2,1)$ irreducible representation
with the $OSp(2,1)$ superspin $j/2$ ($X_{\infty}^i\equiv x^i$ and
$\Theta_{\infty}^{\a}\equiv \tha $).
They are normalized so that
\be [X^m,X^n]=
i{\rho\over\sqrt{{j\over 2}({j\over 2}+{1\over 2})}}\epsilon_{mnp}X^p ,
\qquad\ee
\be [X^i,\Theta^{\a}]=
{\rho\over2\sqrt{{j\over 2}({j\over 2}+{1\over 2})}}{\sigma^i}^{\b\a}
\Theta^{\b},\ee
\be \{\Theta^{\a},\Theta^{\b}\}={\rho\over2\sqrt{{j\over 2}({j\over 2}+
{1\over 2})}}
(C\sigma^i)^{\a\b}X^i,\ee
Hence
\be (X_j^i,X_j^i)_j=(\Theta_j^{\a},\Theta_j^{\a})_j=\rho^2.\ee
$s_{j,lpq}$ are (real) normalization coefficients given by the requirement
that the embedding is norm-conserving.
 Because the operators ${\cal Q}^2,{\cal R}_i^2$ and ${\cal R}_3$ are
hermitian for arbitrary  ${{\cal{SA}}_j}$
(as it can be easily seen from the definitions of the inner products (44),(97))
the embeddings are in fact isometric.
Indeed, the inner product of the eigenvectors
of hermitian operators vanishes if the corresponding
eigenvalues are different. The
commutative limit of the associative product is more
involved, however. We proceed in an analogous way as in the purely bosonic
case \su .

Consider more closely the behaviour of the product as the function
of $k$. According the relation (110),
 arbitrary two elements $\Phi_1,\Phi_2$ of ${{\cal{SA}}_j}$  can be canonically
considered as the elements of ${{\cal{SA}}_k}$ for whatever $k>l$ (including
$k=\infty$). Their product in every  ${{\cal{SA}}_k}$ can also be embedded
in ${\cal{SA}}_{\infty}$. Denote the corresponding element of
${\cal{SA}}_{\infty}$ as $(\Phi_1\Phi_2)_k$.
 We shall argue that
\be \displaystyle{\lim_{k\to\infty}{(\Phi_1\Phi_2)_k}}=\Phi_1\Phi_2\qquad\ee
where $\Phi_1\Phi_2$ is the standard supercommutative pointwise
multiplication in
${\cal{SA}}_{\infty}$ .

For proving the relation (117), it is convenient to realize that
${\cal SA}_j$
can be generated by taking products of generators $X_j^i$ and $\Theta_j^{\a}$
of $OSp(2,1)$ \ in the
irreducible representation with the $OSp(2,1)$ superspin $j/2$. This
statement follows from the Burnside lemma \cite{Nai}, but
its validity can be seen directly. Indeed,  from the \os2 ~commutation
relations it follows easily that every element of the form (111) can be
expressed
in terms of $X_j^i$ and $\Theta_j^{\a}$.
Hence the relations (113-115) ensure the (graded) commutativity in the limit
$j\to\infty$ and it is therefore sufficient just to show that the normalization
coefficients $s_{j,lpq}$ defined in (112) have the property
\be \displaystyle{\lim_{k\to\infty}{s_{k,lpq}}}=s_{\infty,lpq}.\ee
Because of the \os2 ~invariance of the inner products in all ${\cal SA}_k ~
(k=1,\dots,\infty)$, it is in fact enough to
demonstrate it just for the highest
weight elements ${X_k^+}^l$. Then it is a straighforward computation to
check that
\be \displaystyle{\lim_{k\to\infty}}{s_{k,l00}^{-2}}\equiv
\displaystyle{\lim_{k\to\infty}}{({X_k^+}^l,{X_k^+}^l)_k}=
(2l+1)c_{\infty,l0}^{-2},\ee
where $c_{\infty,l0}^{-2}$ have been given in Eq.(17).
But $s_{\infty,l00}^{-2}$ can be directly computed from (44) giving
\be s_{\infty,l00}^{-2}=(2l+1)c_{\infty,l0}^{-2}.\ee
We have thus proven the commutative limit relation (117).

 Note that
  the normalization of $X_j^i$ and $\Theta_j^{\a}$ is such  that the
value of the
Casimir in ${j\over 2}$ $OSp(2,1)$ irreducible representation is equal
to $\rho^2$, i.e.
 \be {X_j^i}^2+C_{\a\b}\Tha_j\Thb_j=\rho^2.\ee
Thus the relation defining the supersphere is preserved also in the truncated
case.
We observe from Eqs. (47) and (116)
that for every $j$
$X_j^i,\Theta_j^{\a}\in {{\cal{SA}}_j}$ are embedded in ${\cal SA}_{\infty}$
as just the standard (super)commutative generators $x^i,\tha$
and
in ${{\cal{SA}}_k},k>j$ as $X_k^i,\Theta_k^{\a}\in {{\cal{SA}}_k}$ .
The notation is therefore
justified and in what follows we shall often write just $X^i$ and
$\Theta^{\a}$.
\subsection{\bf Dirac operator on the truncated sphere}
In an analogy with the (super)commutative case, we  define the non-commutative
spinor bundle on the sphere $S_2$ as the odd part of the truncated superfield
$\Phi\in{\cal{SA}}_j$ and the Dirac operator we define as
\be\rho D\equiv 2({\cal V}_+{\cal V}_- -{\cal V_-}{\cal V_+})-{1\over 2}.\ee
This operator is manifestly self-adjoint, \su ~invariant and it is also odd
with respect to
the grading $\Gamma$ given by Eqs. (93) and (83) or simply, if the diagonal
part of a superfield vanishes,  by
\be \Gamma\Phi_{fer}=\left(\matrix{Id&0\cr 0& -Id}\right)\Phi_{fer}.\ee
This explains the notation in Eq. (92): in the first (second) line there are
right (left)
objects  with respect to the chiral grading $\Gamma$.
Hence, a fermionic  superfield of the upper(lower)-triangular form will be
referred
to as the right (left) chiral spinor on the truncated sphere.

The spectrum of $D$ readily follows from the group representation
considerations. Consider a normalized spinor $\Theta^+/\rho$. It follows
directly from $OSp(2,1)$
graded commutation relations (56-58) that this is the eigenvector of $D$
with an
eigenvalue 1. Moreover it is the highest weight state of one of the
\su ~spin 1/2 representations in the decomposition (82). This can be directly
checked using the generators (93-95):
\be {\cal R_+}\Theta^+=0,\qquad {\cal R}_i^2=3/4.\ee

The construction of the other
(normalized) highest weight states in the irreducible
representations with the higher spins is obvious. They are given by
\be \Psi_{l,h.w.}=
b_{jl}~\rho^{-l-1}\sqrt{{(2l+1)!!\over (2l)!!}} {X^+}^l\Theta^+.
\ee
Here  $l$
is the spin of the \su ~irreducible representation and $b_{jl}$ is a
normalization coefficient. A direct computation shows

\be D\Psi_{l,h.w.}=(l+1)\Psi_{l,h.w.}, \qquad l\leq j-1.\ee
Due to the rotational invariance of $D$ the other eigenvectors within
the irreducible representation are obtained by the action of the lowering
generator ${\cal R}_-$, i.e.
\be \Psi_{l,m}=b_{jl}~\rho^{-l-1}\sqrt{{(2l+1-m)!\over (2l+1)!m!}
{(2l+1)!!\over (2l)!!}}{\cal R}_-^m {X^+}^l \Theta^+.
\ee
The eigenvalue corresponding to the eigenvector $\Psi_{l,m},~~m=0,\dots,
2l$ is obviously $l+1$.
So far we have constructed only one branch
of the spectrum. However, due to an obvious relation
\be D\Gamma+\Gamma D=0\ee
also spinors $\Gamma\Psi_{l,m}$ are the eigenvectors of $D$ with
the eigenvalues $-(l+1)$. In this way we found the complete spectrum because
all eigenvectors $\Psi_{l,m}$ and $\Gamma\Psi_{l,m}$ form the basis of the
space of the fermionic superfields from ${\cal SA}_j$.
Thus, we have obtained precisely the truncation of the commutative Dirac
operator $D$.
\section{\bf Supersymmetric field theories}
\subsection{\bf The bosonic preliminaries}

Consider the following action for a real scalar field living on the
sphere $S_2$
\be S(\phi)={1\over 2}(\phi,R_i^2\phi)_{\infty}\equiv{1\over 4\pi\rho}\int
d^3 x^i\delta({x^i}^2-\rho^2)\phi(x)R_i^2\phi(x).\ee
It is easy to show that this is just the action of a free massless
field on $S_2$ i.e.
\be S(\phi )=-{1\over 8\pi}\int d\Omega\phi\bigtriangleup_{\Omega}\phi,\ee
where $\bigtriangleup_{\Omega}$ is the Laplace-Beltrami operator on
the sphere or, simply, the angular part of the flat Laplacian in $R^3$.
Adding a mass and an interaction term is easy, e.g. the
$P(\phi )$-models \cite{GJ,GKP1} are described by the action
\be S_{\infty}={1\over 2}(\phi,R_i^2\phi)_{\infty}+(1,P(\phi))_{\infty},\ee
where $P(\phi )$ is a polynomial in the field variable. The
non-commutative analogue of the action (129) is now obvious
\be S_j ={1\over 2}(\phi,{\cal R}_i^2\phi)_j+(1,P(\phi))_j=
                    {1\over 2j+2}Tr_j(\phi{\cal R}_i^2\phi)+{1\over j+1}Tr_j
      P(\phi).\ee
The truncated action is manifestly \su invariant with respect to
the infinitesimal transformation of the scalar field
\be \delta\phi=\e_i{\cal R}_i\phi\equiv\e_i[R_i,\phi].\ee

Another interesting class of Lagrangians consists of the nonlinear
$\sigma$-models describing the string propagation in curved
backgrounds. The (truncated) action reads
\be S_j={1\over 2}({\cal R}_i\phi^A,g^{AB}(\phi){\cal R}_i\phi^B)_j\ee
with the obvious commutative limit.
It is not difficult, in fact, to define a quantization of the
truncated system via the path integral because the space of field
configurations in finite-dimensional. We gave the details in
a separate publication \cite{GKP1} with the aim to develop the efficient
nonperturbative regularization of field theories which could
 (hopefully in many
aspects) compete with the traditional lattice approach.
\subsection{\bf The supersymmetric actions}
The supersymmetric case is somewhat more involved than the bosonic
one not only because of the enlargement of the number of degrees of
freedom. Starting from the undeformed case one could suspect that the
standard free $OSp(2,1)$-supersymmetric action for a real superfield on the
sphere should be written in our three dimensional formalism as
\be S_{susp}={1\over 2}(\Phi,({\cal R}_i^2+
C_{\a\b}{\cal V}_{\a}{\cal V}_{\b})\Phi)_{\infty}.
\ee
Though the $OSp(2,1)$  Casimir sitting within the brackets does give the
SUSY invariance it {\it does not} yield the correct two dimensional
"world-sheet" action containing just the free massless bosonic field
and free massless Majorana fermion. To get out of the trouble we may
use the philosophy used about a decade ago where supersymmetric
models on the homogeneous spaces have been intensively studied
\cite{Fron}.
In particular, Fronsdal has considered the spinors on anti-de Sitter
spacetime and has constructed the $OSP(4,1)$ invariant supersymmetric
actions by introducing another set of odd generators \cite{Fron}. They  were
analogues of the standard supersymmetric covariant derivatives needed
to build up the super-Poincar\'e invariant Lagrangians.

The same approach applies in our case. The new odd generators
are nothing but the additional \os2 ~ generators ${\cal D}_{\a}$. The standard
Lagrangian of the free $OSp(2,1)$ supersymmetric theory can be written
solely in
terms of the `covariant derivatives' ${\cal D}_{\a}$ and the grading $\G$.

Let us begin with the detailed quantitative account first in the
non-deformed case.
It is easy to check that the operator
\be C_{\a\b}d_{\a}d_{\b}+{1\over 4}\G_{\infty}^2\ee
is invariant with respect to $OSp(2,1)$ supersymmetry generated by $r_i$
and $v_{\pm}$. Hence we may consider the action
$$ S=(\Phi,C_{\a\b}d_{\a}d_{\b}\Phi)_{\infty}+
{1\over 4}(\Phi,\G_{\infty}^2\Phi)_{\infty}\equiv $$
$$\equiv{\rho\over 2\pi}\int_{R^3}d^3 x^i d\theta^+ d\th-
\delta({x^i}^2+C_{\a\b}\tha\thb -\rho^2)\Phi(x^i,\tha)
(\C d_{\a}d_{\b} +{1\over 4}\G^2)\Phi(x^i,\tha),\eqno(136a)$$
where $\Phi$ is a real superfield, i.e. $\Phi^{\dda}=\Phi$.

Consider now the variation of the real superfield $\Phi$
\be \delta\Phi=i\e_{\a}v_{\a}\Phi,\ee
which preserves the reality condition. Now Eqs.(63-66) hold also when
$\Phi_1$ is an even and $\Phi_2$ an odd superfield in the standard Grassmann
sense. Using this  and the fact
that $\e_{\a}v_{\a}$ commutes with the operator (136), the
supersymmetry of the action $S$ obviously follows.

It is straightforward to work out the action (136a) in the
two-dimensional component language. It reads
\be S={1\over 4\pi}\int d\Omega (-{1\over 2}\phi\bigtriangleup_{\Omega}\phi+
{1\over 2}
\rho^4 F^2 -{1\over 2}\psi^{\da}\rho^3 D_{\Omega}\psi),\ee
where $D_{\Omega}$ is the Dirac operator on $S^2$ and the superfield
ansatz is
\be \Phi(x^i,\tha)=\phi(x^i)+\psi_{\a}\tha +(F+{x^i\over r^2}\del_i
\phi)\theta^+ \th- .\ee
Of course, $\psi_{\a}$ are anticommuting objects and the reality
condition $\Phi^{\dda}=\Phi$ makes the fields $\phi$ and $F$ real
and the spinor $\psi_{\a}$ becomes Majorana\footnote{Note, that we
consider the {\it graded} involution defined by Eq.(45) (see also
\cite{SNR2}).}, i.e.
\be \psi_+^{\dda}=\psi_-,\qquad \psi_-^{\dda}=-\psi_+.\ee
We recognize in the expression (138)
the standard free supersymmetric action in two
dimensions.

Adding a (real) superpotential $W(\Phi)$ we may write down a
supersymmetric action with the interaction term. It reads
\be S_{\infty}=(\Phi,(\C d_{\a}d_{\b}+{1\over 4}\G_{\infty}^2)\Phi)_{\infty}+
(1,W(\Phi))_{\infty}.\ee

 The truncated version of the action $S_{\infty}$
\be S_j=(\Phi,(\C {\cal D}_{\a}{\cal D}_{\b}+{1\over 4}\G^2)\Phi)_j+
(1,W(\Phi))_j\ee
is manifestly supersymmetric with respect to the variations
\be \delta\Phi=i\epsilon_{\a}{\cal V}_{\a}\Phi,\ee
It remains to prove that $S_j$ approaches $S_{\infty}$ for
$j\to\infty$. In order to do that it is convenient to rewrite
both truncated and untruncated action as follows
\be S_j=({\cal D}_+\Phi,{\cal D}_+\Phi)_j
+({\cal D}_-\Phi,{\cal D}_-\Phi)_j+{1\over 4}(\G\Phi,\G\Phi)_j+
(1,W(\Phi))_j,\ee
where the index j can be both finite and infinite and we have used
the formulas (63-66) and (101-104). Now it is enough to show that
\be \displaystyle{\lim_{k\to\infty}{({\cal D}_{\a}\Phi)_k}}=d_{\a}\Phi,
\qquad
\displaystyle{\lim_{k\to\infty}{(\G\Phi)_k}}=\G_{\infty}\Phi\ee

(The embedding $(\Phi)_k$ was defined in Eqs.(111,112).) But this is true
almost by definition because ${\cal D}_{\a} \Phi$ can be written as a linear
superposition
of the vectors of the form (111,112).
As in the bosonic case we may write down the regularized action
for the supersymmetric $\sigma$-models describing the superstring
propagation in curved backgrounds

\be S_j=({\cal D}_+\Phi^A,g_{AB}(\Phi){\cal D}_+\Phi^B)_j
+({\cal D}_-\Phi^A,g_{AB}(\Phi){\cal D}_-\Phi^B)_j+{1\over 4}(\G\Phi^A,
g_{AB}(\Phi)\G\Phi^B)_j.\ee
The $OSp(2,1)$ supersymmetry and the commutative limit is obvious.
 The regularized action (146) can
be used as the base for the path integral quantization manifestly
preserving supersymmetry and still involving the finite number of
degrees of freedom. Particularly this aspect of our approach seems
to be very promising both in comparison with the lattice physics
as well as in general. Indeed so far we are not aware of any
nonperturbative regularization which would possess all those
properties.
\section{Conclusions and Outlook}
We have regulated in the manifestly supersymmetric way the actions
of the field theories on the supersphere, involving scalar and spinor
fields. As a next step we plan to include in the picture the
topologically non-trivial bundles and the gauge fields \cite{GKP} and
to study the chiral symmetry in the context. From the purely mathematical
point of view we have to build up the non-commutative de Rham complex
and understand the notions of one- and two-forms. It would be
also interesting to establish a connection between previous works
on supercoherent states \cite{ChEP,Elg,ElgN} and our present treatment.
In a later future we shall attempt to reach two challenging goals in
our programme, namely the truncation of the four-dimensional sphere
and the inclusion of gravity.
\section{Acknowledgement}
We are grateful to A. Alekseev, L.
\'Alvarez-Gaum\'e, M. Bauer, A. Connes,
V. \v Cern\'y, T. Damour, J. Fr\"ohlich, J. Ft\'a\v cnik,
K. Gaw\c edzki, J. Hoppe, B. Jur\v co, E. Kiritsis, C. Kounnas, M. Rieffel,
R. Stora and D. Sullivan
for useful discussions. Part of the research of C.K. has been done
at I.H.E.S. at Bures-sur-Yvette and of C.K. and P.P. at the
Schr\"odinger Institute in Vienna. We thank both these institutes
for hospitality.

\end{document}